\documentclass[12pt]{article}
\usepackage{graphicx}
\usepackage{authblk}
\usepackage{color}
\usepackage[small,labelfont=bf,up]{caption}

\newcommand\pubnumber{DPF2015-350}
\newcommand\pubdate{\today}

\def\napoli{Mu2e Collaboration Cosmic Ray Veto Group}
\def\JINR{Joint Institute for Nuclear Research}
\def\UVA{University of Virginia}
\def\NIU{Northern Illinois University}
\def\FNAL{Fermi National Accelerator Laboratory}

\def\Title#1{\begin{center} {\bf\Large #1 } \end{center}}

\def\Address#1{\begin{center}{\textbf{\it #1}} \end{center}}

\newcommand\pubblock{\rightline{\begin{tabular}[!htbp]{l} \pubnumber\\
         \pubdate  \end{tabular}}}
\newenvironment{Abstract}{\begin{quotation}  }{\end{quotation}}
\newenvironment{Presented}{\begin{quotation}\begin{center}
\hspace{-5mm}PRESENTED AT\end{center}\bigskip
      \begin{center}\begin{large}}{\end{large}\end{center} \end{quotation}}

\def\beq{\begin{equation}}
\def\eeq#1{\label{#1}\end{equation}}
\def\eeqn{\end{equation}}

\def\beqa{\begin{eqnarray}}
\def\eeqa#1{\label{#1}\end{eqnarray}}
\def\eeqan{\end{eqnarray}}

\let\bar=\overbar

\def\Dslash{\not{\hbox{\kern-4pt $D$}}}
\def\dslash{\not{\hbox{\kern-2pt $\del$}}}

\def\msb{{\bar{\ssstyle M \kern -1pt S}}}

\title{\Title{Performance of Scintillator Counters with Silicon
Photomultiplier Readout}}

\author[$\ast$]{A.~Artikov}
\author[$\ast$]{V.~Baranov}
\author[$\ast$]{D.~Chokheli}
\author[$\ast$]{Yu.~I.~Davydov}
\author[$\dag$]{E.~C.~Dukes}
\author[$\dag$]{R.~Ehrlich}
\author[$\ddag$]{K.~Francis}
\author[$\dag$]{M.~J.~Frank}
\author[$\ast$]{V.~Glagolev}
\author[$\dag$]{R.~C.~Group}
\author[$\S$]{S.~Hansen}
\author[$\S$]{A.~Hocker}
\author[$\dag$]{Y.~Oksuzian}
\author[$\S$]{P.~Rubinov}
\author[$\dag$]{E.~Song}
\author[$\ddag$]{S.~Uzunyan}
\author[$\dag$]{Y.~Wu}
\affil[$\ast$]{\JINR}
\affil[$\dag$]{\UVA}
\affil[$\ddag$]{\NIU}
\affil[$\S$]{\FNAL}

\date{}

\begin{document}
\begin{titlepage}
\pubblock
\vspace{-5mm}
{\let\newpage\relax\maketitle}
\thispagestyle{empty}
\vspace{-7mm}
\Address{\napoli}
\vspace{7mm}
\begin{Abstract}
The performance of scintillator counters with embedded wavelength-shifting
fibers has been measured in the Fermilab Meson Test Beam Facility using
$120~\mathrm{GeV}$ protons. The counters were extruded with a titanium dioxide
surface coating and two channels for fibers at the Fermilab NICADD facility.
Each fiber end is read out by a $2\times2~\mathrm{mm^2}$ silicon
photomultiplier. The signals were amplified and digitized by a custom-made
front-end electronics board. Combinations of $5\times2~\mathrm{cm^2}$ and
$6\times2~\mathrm{cm^2}$ extrusion profiles with $1.4$ and
$1.8~\mathrm{mm}$ diameter fibers were tested. The design is intended for the
cosmic-ray veto detector for the Mu2e experiment at Fermilab. The light yield as
a function of the transverse and longitudinal position of the beam will
be given.
\end{Abstract}
\vspace {5mm}
\begin{Presented}
DPF 2015\\
The Meeting of the American Physical Society\\
Division of Particles and Fields\\
Ann Arbor, Michigan, August 4--8, 2015\\
\end{Presented}
\vfill
\end{titlepage}
\def\thefootnote{\fnsymbol{footnote}}
\setcounter{footnote}{0}
\section{Introduction}
The Mu2e experiment at Fermilab intends to make the most sensitive measurement
of the neutrinoless, coherent conversion of muons into electrons in the field of
a nucleus~\index{tdr}\cite{tdr}:
\[
\mu+N \to e+N
\]
This process is an example of charged lepton flavor violation, and its detection
would be unambiguous evidence for physics beyond the Standard Model.

\begin{figure}[h!]
\centering
\vspace{5mm}
\includegraphics[width=5in]{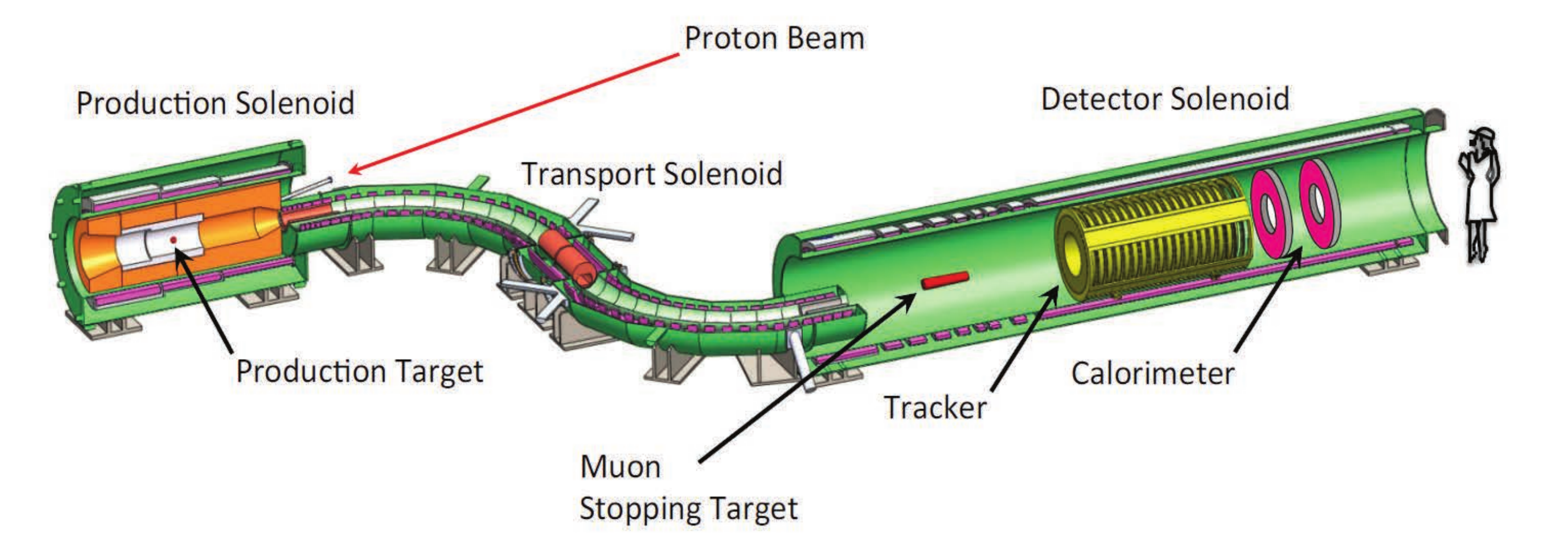}
\caption{The layout of the Mu2e experiment~\cite{tdr}.}
\label{fig:mu2elayout}
\end{figure}

The layout of the Mu2e experiment is shown in
Figure~\ref{fig:mu2elayout}~\cite{tdr}. The high-intensity pulsed
proton beam impacting the production target produces muons among other
particles. The muons propagate through the transport solenoid and roughly half
are captured in the stopping target.  If a muon is converted into an electron
via the conversion process, a $105~\mathrm{MeV}$ electron would be ejected
from the stopping target, which should be detected by the tracker and calorimeter.

One major background source for the Mu2e experiment is cosmic ray-induced muons
faking signal electrons by interacting with the detector materials~\cite{tdr}.
Simulations show such background events occur at a rate of approximately one per
day. For Mu2e to be able to reach its sensitivity goal, this background must
be suppressed to $\sim$0.1 event over the entire 3-year lifetime of the
experiment~\cite{tdr}. In order to combat this background, Mu2e relies on a
cosmic ray veto~(CRV) system.

\begin{figure}[h!]
\centering
\includegraphics[width=4.6in]{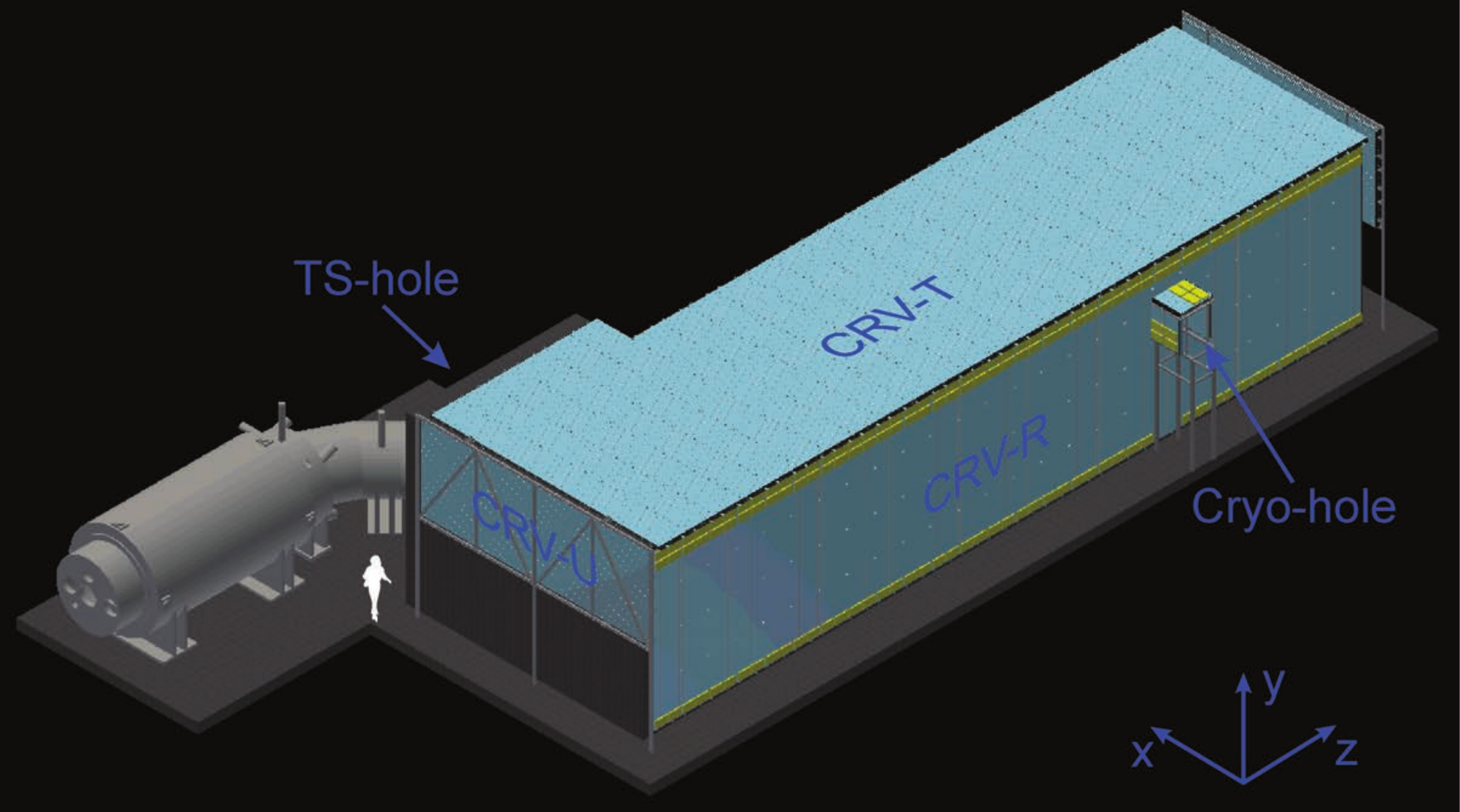}
\caption{A schematic of the cosmic ray veto (CRV) detector. It covers the detector
solenoid and part of the transport solenoid shown in
Figure~\ref{fig:mu2elayout}, but there is no coverage underneath the detector.}
\label{fig:crv1}
\end{figure}

\begin{figure}[h!]
\centering
\includegraphics[width=4.7in]{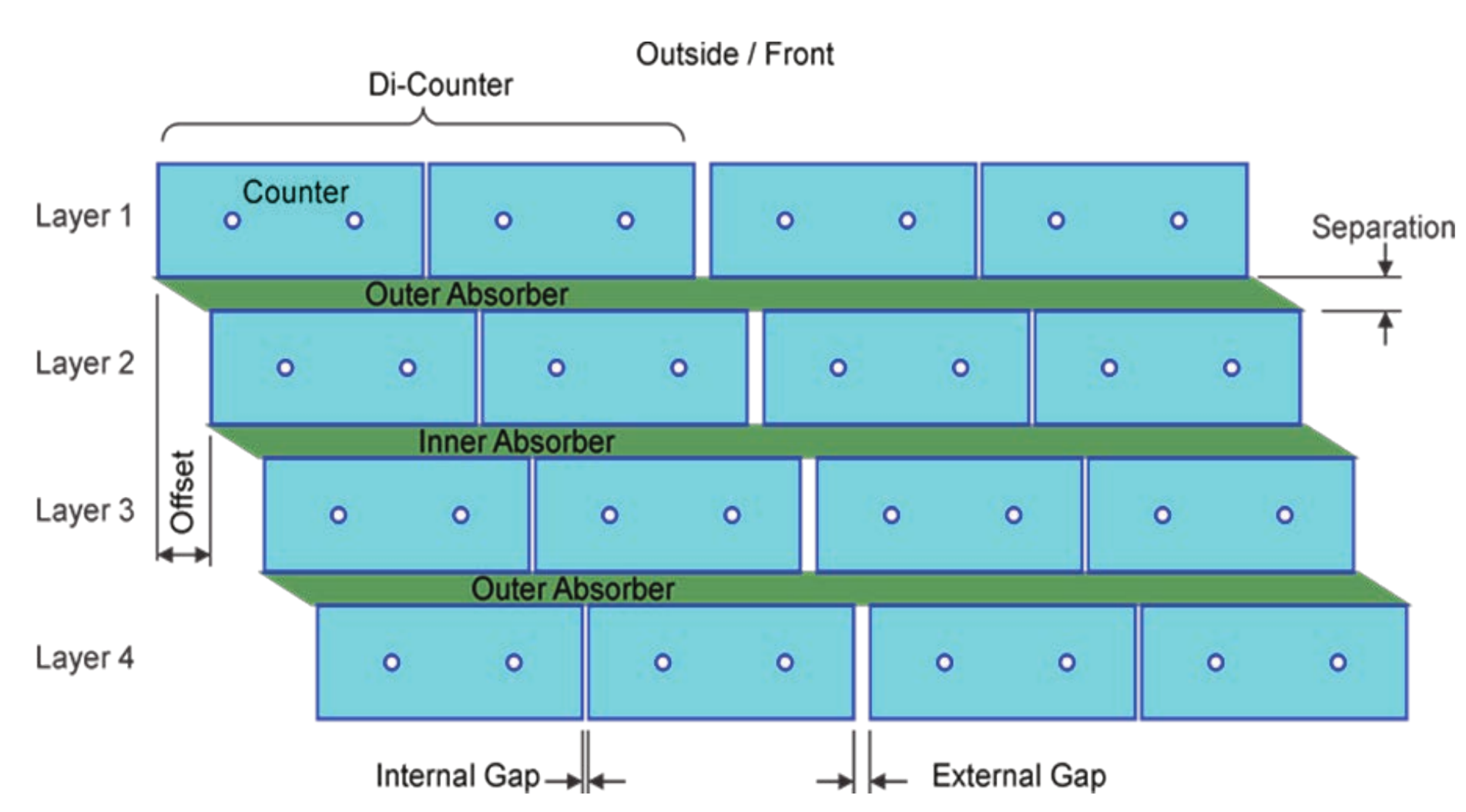}
\caption{A schematic drawing of the cross-sectional view of the CRV counters.
One rectangular box corresponds to a scintillator counter with a thickness of
$20~\mathrm{mm}$, a width of 50 or $60~\mathrm{mm}$, and a length of up to
$6,600~\mathrm{mm}$. Each counter contains two holes for the wavelength-shifting
fibers which propagate the light signals to the silicon photomultipliers on both
ends\protect\footnotemark. Two scintillator counters are glued together to form one di-counter.}
\label{fig:crv2}
\end{figure}

Figure~\ref{fig:crv1} shows a schematic view of the Mu2e CRV
detector~\cite{tdr}.
The CRV covers the majority of the detector solenoid and about half of the
transport solenoid shown in Figure~\ref{fig:mu2elayout}~\cite{tdr}. Simulations
have shown that the vetoing efficiency of the CRV has to be at least $99.99\%$
and must withstand an intense radiation environment. Each side of the CRV
consists of four staggered layers of scintillator counters
(Figure~\ref{fig:crv2}). Each scintillator counter contains two embedded
wavelength shifting fibers.

The scintillator counters were tested at Fermilab in a $120~\mathrm{GeV}$ proton
beam~\index{fermi}\cite{fermi}. The goal of the test was to measure the lateral and longitudinal response
of the scintillator counter at different beam angles of incidence with 50 and
60-mm wide counters containing 1.4 and 1.8-mm diameter fibers.

\footnotetext{A small fraction of the counters in the detector are read out only from one end yet
all the counters tested have readouts on both ends.}

\section{Test Beam Setup}
\begin{figure}[h!]
\centering
\includegraphics[width=5in]{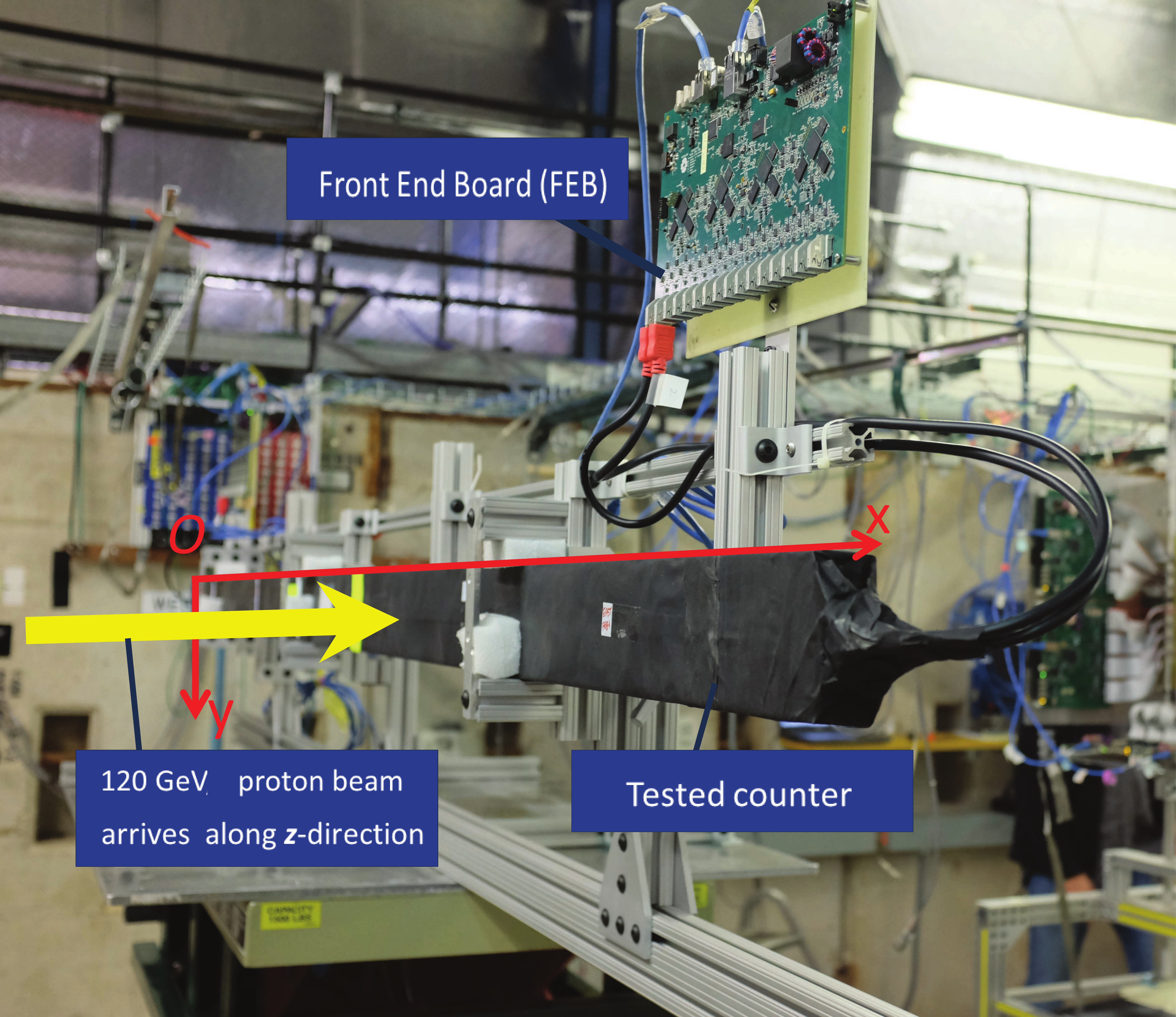}
\caption{A photograph of the test beam setup. The coordinate system is indicated
in red showing the direction of the $x$- and $y$-axes together with the origin. 
The yellow arrow indicates the proton beam direction.}
\label{fig:setup}
\end{figure}

\begin{figure}[h!]
\centering
\includegraphics[width=4in]{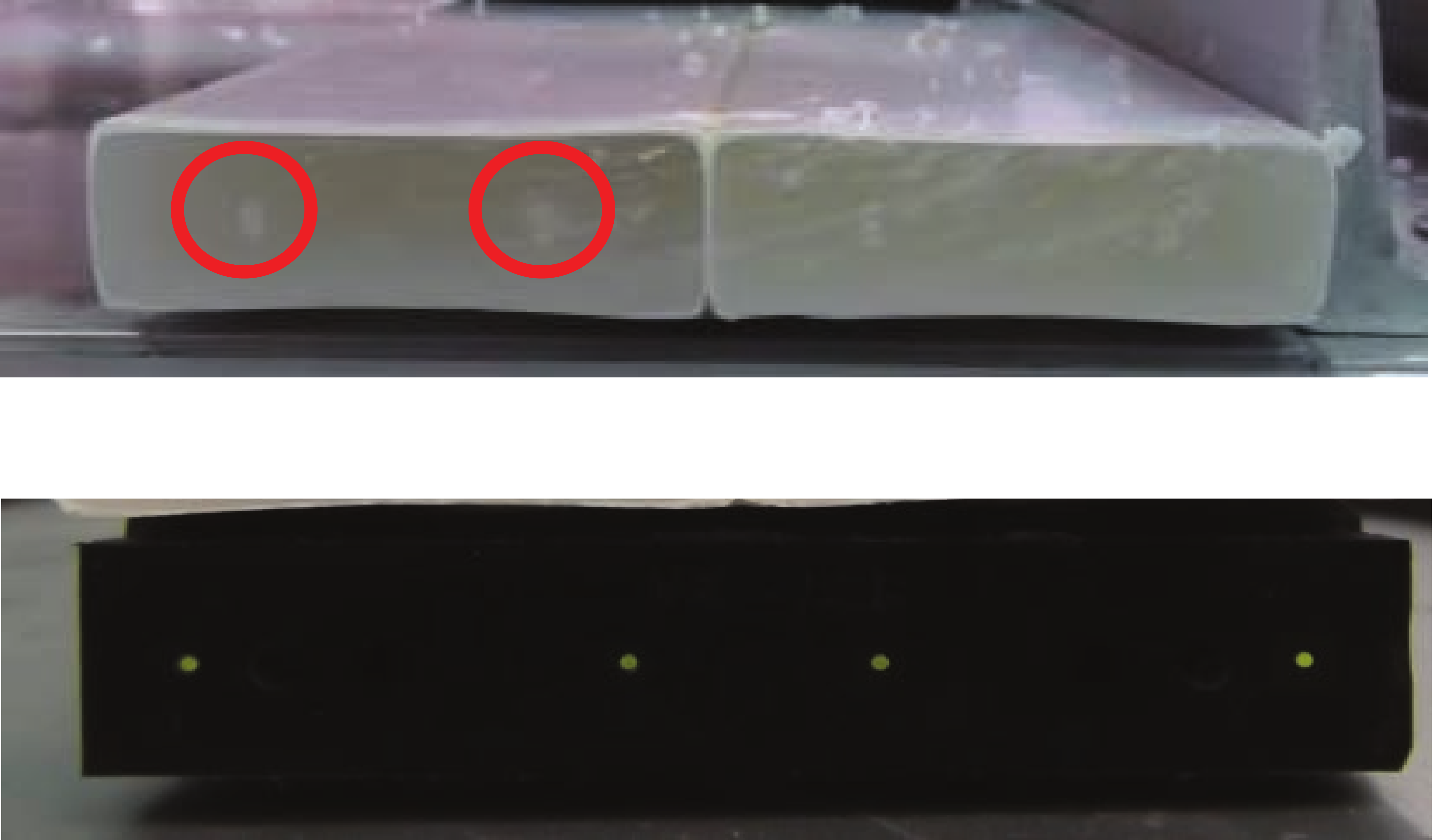}
\caption{Top: Two scintillator counters are glued together to
form one di-counter. The extrusions are coated with titanium dioxide so
light cannot travel from one counter to the other. Two extrusion sizes were
tested: $20~\mathrm{mm}\times50\mathrm{mm}\times3000~\mathrm{mm}$ and
$20~\mathrm{mm}\times60\mathrm{mm}\times1820~\mathrm{mm}$. Each extrusion is
equipped with two wavelength-shifting fibers as indicated by the red circles. The
fibers (Kuraray Y11 non-s type, 175 ppm) are not glued in their channels. Two
fiber diameters were tested: $1.4~\mathrm{mm}$ and $1.8~\mathrm{mm}$.
Bottom: Fibers are glued in place to the fiber guide bar and polished using a
diamond flycutter.}
\label{fig:crosssection}
\end{figure}

\begin{figure}[h!]
\centering
\includegraphics[width=2.2in]{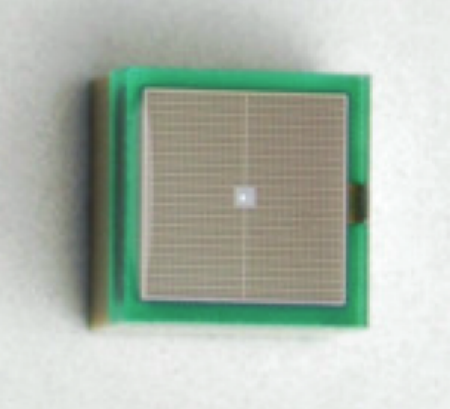}
\caption{Photograph of a Silicon Photo-Multiplier (SiPM) used to read out the fibers.  We use the
Hamamatsu S13360-2050VE SiPM with a photosensitive area of
$2~\mathrm{mm}\times2~\mathrm{mm}$ and a pixel pitch of $50~\mathrm{\mu m}$.
Each SiPM contains approximately 1,600 pixels.}
\label{fig:sipm}
\end{figure}

\begin{figure}[h!]
\centering
\includegraphics[width=4.5in]{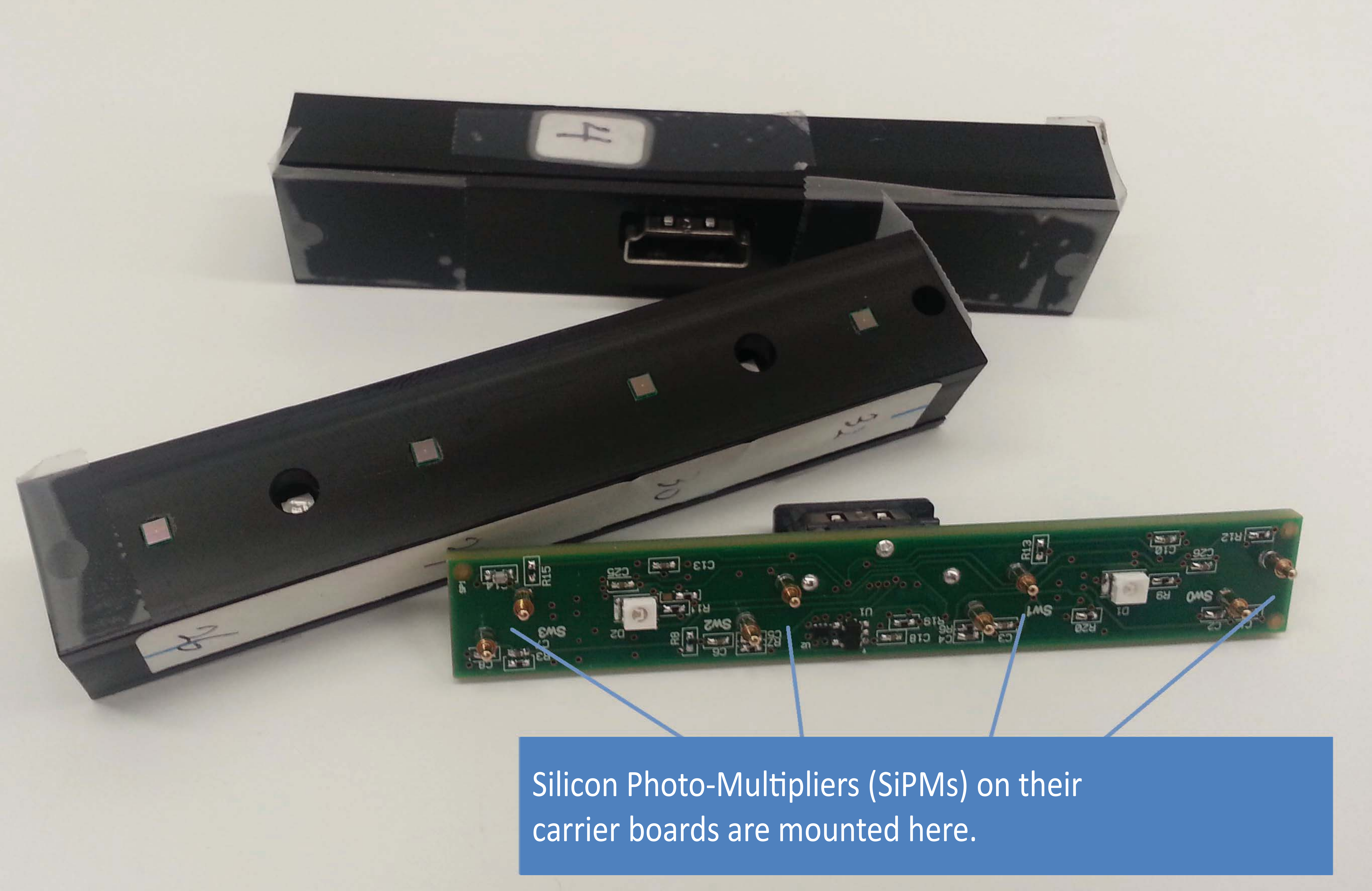}
\caption{The counter motherboard (CMB) has two LED
flashers, a thermometer and spring-loaded pins that push the SiPM carrier boards
(not shown) against the fibers. One CMB is mounted on each end of a di-counter,
so there is a total of two CMBs per di-counter.}
\label{fig:cmb}
\end{figure}

Figure~\ref{fig:setup} shows a photograph of the test beam setup. The
scintillator counters were mounted on the test stand with the coordinate system
shown in Figure~\ref{fig:setup}. The 120~GeV protons are minimum ionizing, so
they deposit a similar energy as minimum-ionizing muons. When a
charged particle traverses the counter and deposits energy in the
scintillator, light is produced which is captured by the
wavelength-shifting fibers shown in~Figure~\ref{fig:crosssection}.
The light is retransmitted to the ends of the counter where it is detected and
amplified by Silicon Photo-Multipliers (SiPMs) shown in
Figure~\ref{fig:sipm}. The SiPMs are mounted on small carrier boards that sit
in wells which allow them to be pushed up against the fibers. A counter
motherboard (CMB) as shown in Figure~\ref{fig:cmb} with spring-loaded pins makes
electrical contact with the SiPM carrier boards. Signals from up to 16 counter
motherboards are sent to a front-end board by HDMI cables where they are
amplified and digitized.

The counters are mounted on a rotatable support fixture on a table capable of 
translational motion, which allows us to place the proton beam at different positions on the
counter and to vary incident angles. The change in incident angle changes the
path length of the protons through the scintillator and therefore
changes the light yield. A coincidence of beam scintillator counters read out
by photomultiplier tubes is used to provide a trigger. Upon receipt of a trigger
signal, the SiPMs were read out for $1.6~\mathrm{\mu s}$ with their signals
digitized every $12.6~\mathrm{ns}$. Four multiwire proportional chambers (MWPC),
two upstream and two downstream of the CRV counters, provide tracking
information~\cite{mwpcbibli}, which allow us to reconstruct the proton
trajectory to within $0.25~\mathrm{mm}$ at the counters.

\section{Data Analysis}
\begin{figure}[h!]
\centering
\includegraphics[width=4.5in]{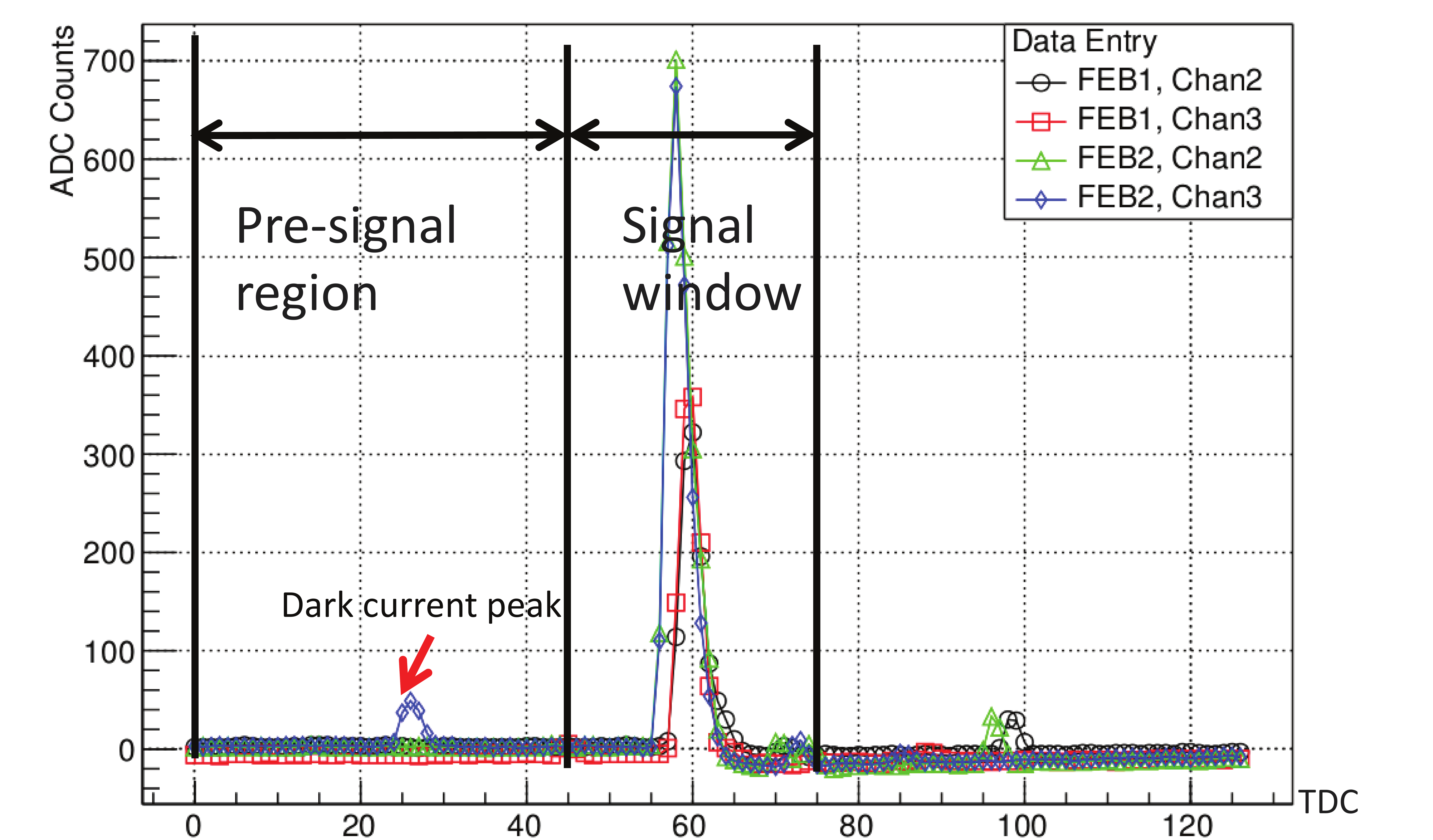}
\caption{A typical event is shown here. Each TDC corresponds to about
$12.6~\mathrm{ns}$. Nominal pedestal has been subtracted. In the figure four channels of a single scintillation
counter are shown: channels 2 and 3 of FEB1 are on the $+\hat x$ end; channels 2
and 3 of FEB2 are on the $-\hat x$ end. The peak in the signal region is caused by a
proton, and the small signal in the pre-signal region is a dark current pulse.}
\label{fig:signal}
\end{figure}

Figure~\ref{fig:signal} shows a typical event with all four channels on an
extrusion. The pulses before the signal window are dark current peaks, which are
used for calibration.

\subsection{Calibration}
\begin{figure}[h!]
\centering
\includegraphics[width=5in]{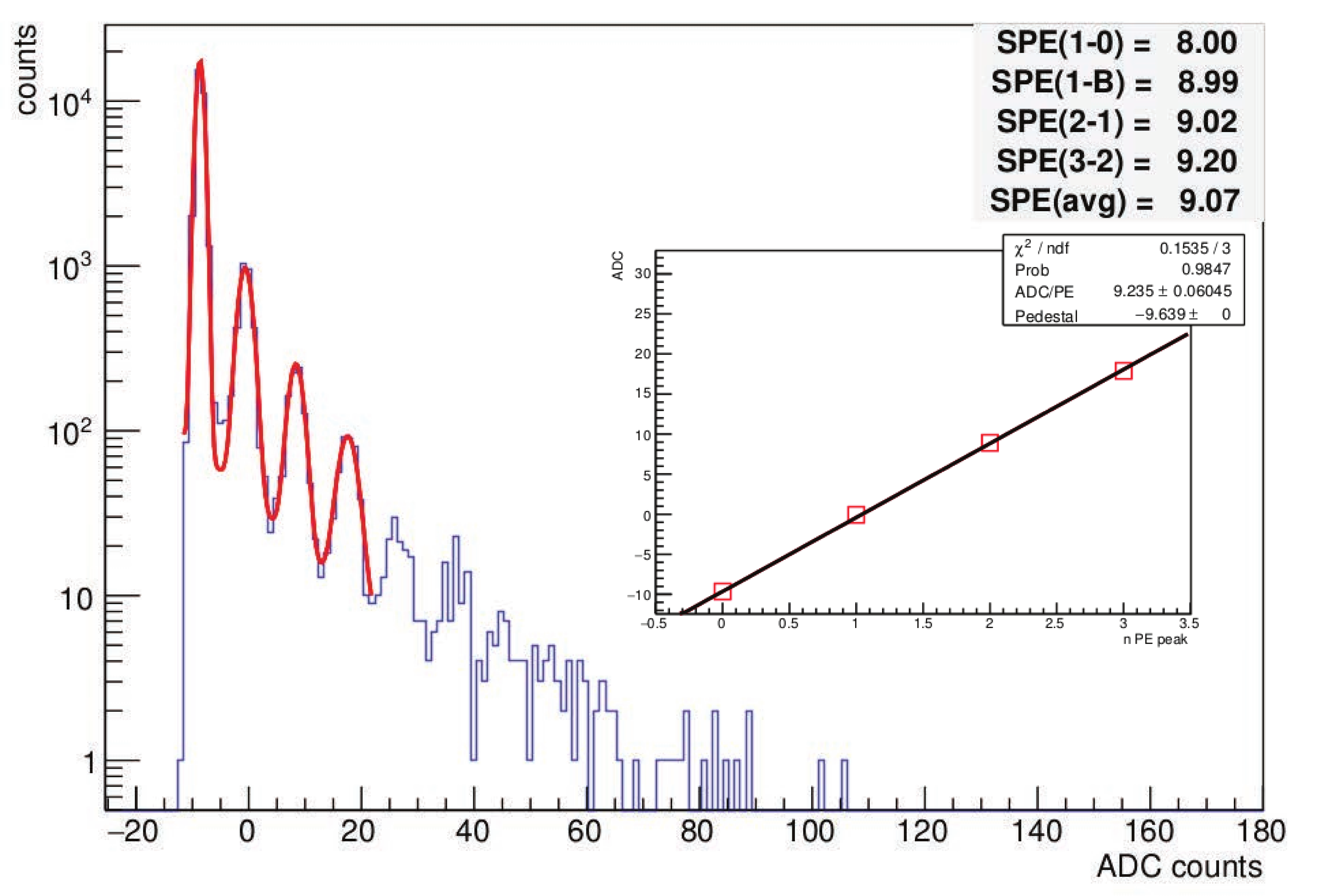}
\caption{The dark current (noise) spectrum of a single SiPM. The peaks
correspond to 0, 1, 2, 3~\ldots pixels that fire in the SiPM.}
\label{fig:darkpeak}
\end{figure}

Since the SiPMs produce resolvable single pixel peaks, a calibration can be
carried out to convert ADC units to the number of pixels. The distribution of
the dark current peaks allows such a conversion. As shown in
Figure~\ref{fig:darkpeak}, by fitting a function to the distribution, the
positions of the peaks can be extracted and the average ADC per pixel fired can
be calculated. The number of pixels fired is proportional to the number of
incident photons with crosstalk. To find the photoelectron yield the crosstalk
needs to be determined as discussed later.

\subsection{Transverse and Longitudinal Response}
After calibration the number of pixels fired in the SiPM for each
event can be calculated. Then the transverse
and longitudinal response of the extrusion can be studied.

\begin{figure}[h!]
\centering
\includegraphics[width=4in]{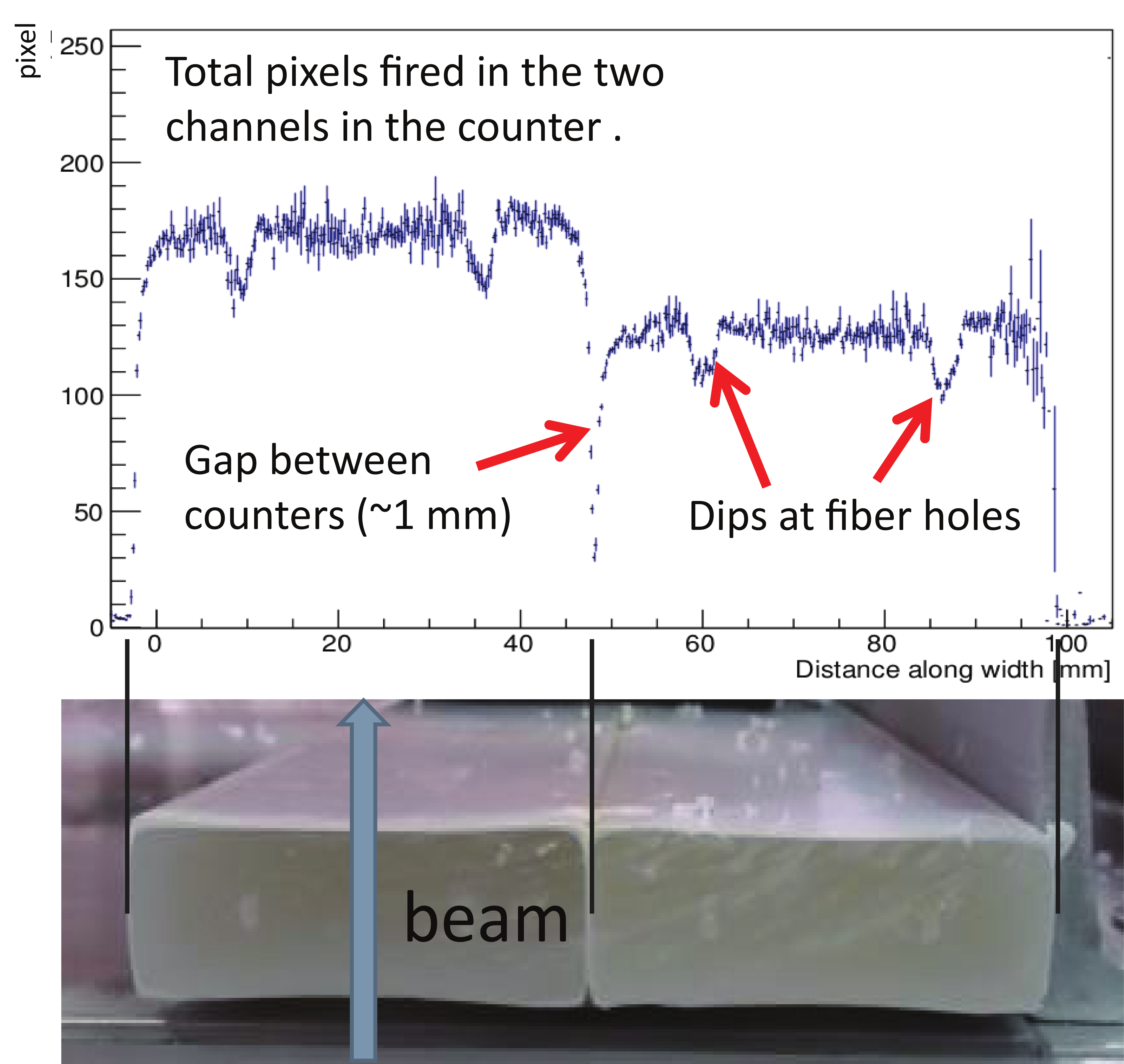}
\caption{Transverse response of a 100-mm wide di-counter. The
illustration shows the number of pixels fired (sum of the two SiPMs on the same
end of the extrusion) as the beam scanned through the width of the di-counter at
$x=1,400~\mathrm{mm}$. Features of the counter, including the gap between the
counters and the fiber holes are visible.}
\label{fig:transverse}
\end{figure}

Figure~\ref{fig:transverse} shows the transverse response with the beam
incident at $x=1,400~\mathrm{mm}$. The total number of pixels fired in the
two channels on the $-\hat x$ end of the extrusion is plotted. From the figure it can
be observed that the number of fired pixels is relatively constant across the
width of the counter. The edges of the counter are well defined.
The signal size does not fall below 50\% until the beam incident position
reaches about $0.5~\mathrm{mm}$ from the edges. The gap between the two
extrusions is clearly evident. It has an effective width of about
$1~\mathrm{mm}$. Dips of 20\% to 25\% in the signal are also present at the
locations where the fiber holes are. Such dips are expected and arise due to
the reduced path length in the scintillator when the proton passes through one
of the fiber holes. The difference between the total number of fired pixels in
the two counters is very likely due to the different crosstalk levels between the
SiPMs, which will be discussed in the following section.

\begin{figure}[h!]
\centering
\includegraphics[width=6.25in]{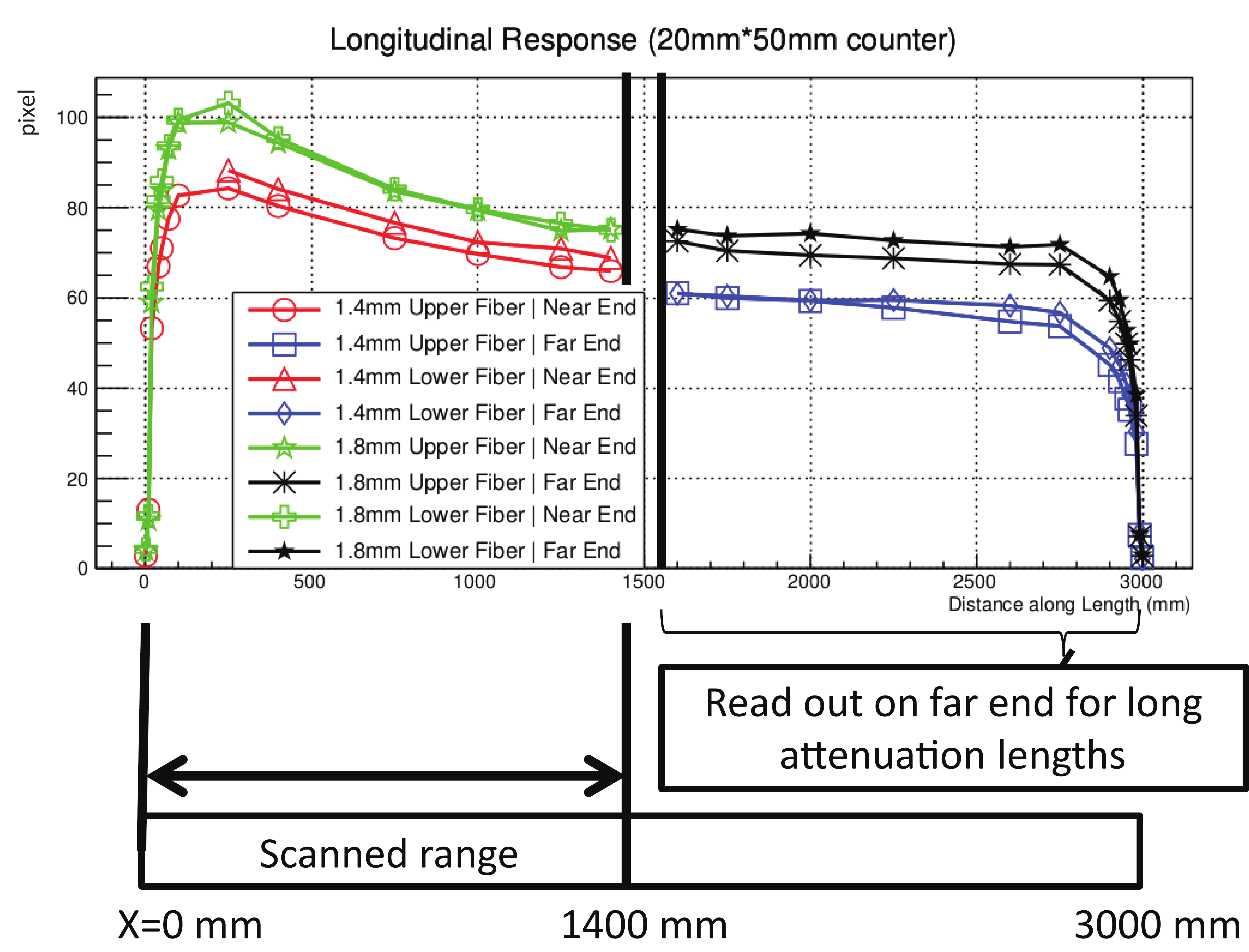}
\caption{Longitudinal response of two counters with different size fibers.}
\label{fig:longitudinal1}
\end{figure}

Figure~\ref{fig:longitudinal1} shows the longitudinal response of the
extrusions with the beam centered between two fibers. Due to the limitation of
the test facility, a full-length scan was impossible; instead, half the length
of the extrusion was scanned ($0\sim1,400~\mathrm{mm}$). In
Figure~\ref{fig:longitudinal1} points greater than $1,400~\mathrm{mm}$ come from
SiPM at the far ($+\hat x$) end. Each data point corresponds to the average
number of fired pixels of a single SiPM. The nice aggreement between the data
taken from different channels indicates a consistency among the SiPMs.
Simulations show the response of the counter along its length is consistent with
the fiber attenuation length. Figure~\ref{fig:longitudinal1} also shows that the
light output of the 1.4-mm and 1.8-mm diameter fibers has a difference of about
15\%.\footnote{Recent bench-test measurements using sources and cosmic ray muons
show a larger increase of almost 40\%, which is expected. Misalignment at the
the $1.8~\mathrm{mm}$ fiber and the $2~\mathrm{mm}$ SiPMs is presumed to be
the reason for the lower increase observed here.}

\begin{figure}[h!]
\centering
\includegraphics[width=6.25in]{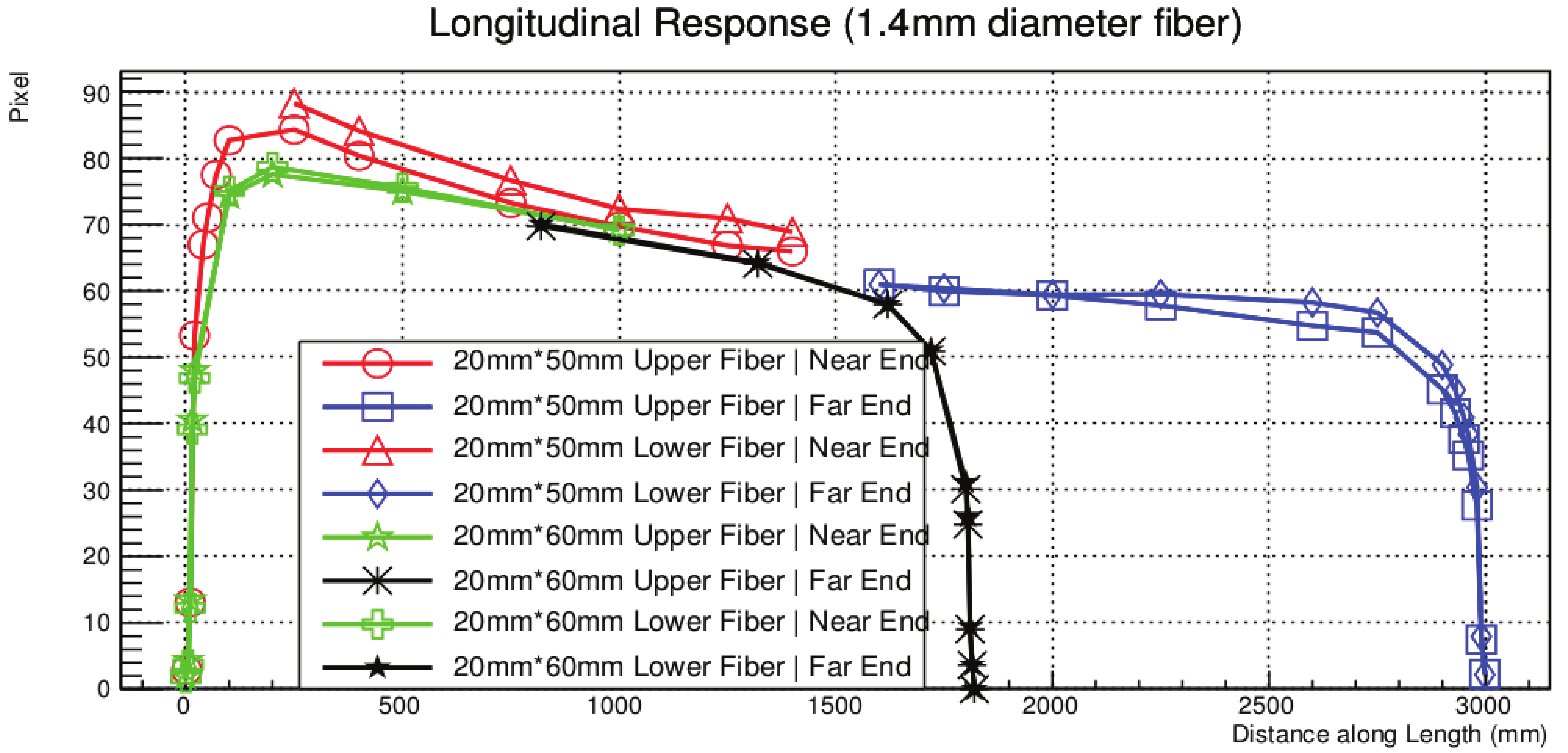}
\caption{Longitudinal response of two counters with different widths, both
with 1.4-mm diameter fibers.}
\label{fig:longitudinal2}
\end{figure}

Figure~\ref{fig:longitudinal2} shows the small performance
difference between the 50-mm and 60-mm wide extrusions.

\begin{figure}[h!]
\centering
\includegraphics[width=6.25in]{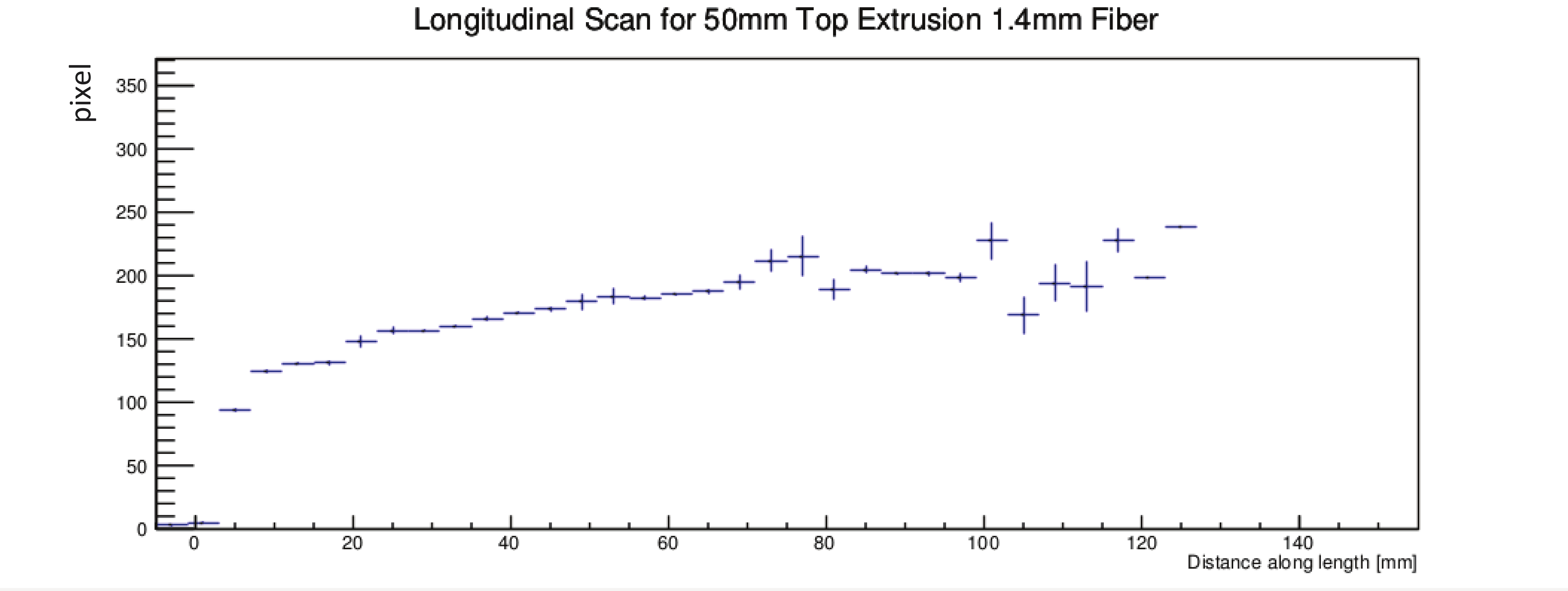}
\caption{Close-up view of the longitudinal response at small x values. The
outputs of both SiPMs at the $-\hat x$ end have been summed.}
\label{fig:longitudinal3}
\end{figure}

Figure~\ref{fig:longitudinal3} provides a
close-up view of the longitudinal response at small x values.A fall-off is
present in the longitudinal response towards the ends, which is
thought to be caused by photons being absorbed by the dark-colored fiber guide
bars. Detailed Monte Carlo studies are underway.

\subsection{Photoelectron Yield and Cross Talk}
We have been careful to only report the pixel yield of the devices. To obtain
the photoelectron yield from the pixel yield, the effect of cross talk needs to
be considered. By crosstalk we mean pixels that induce neighboring pixels to
fire, which inflates the pixel count. Hamamatsu, the SiPM provider, reports a
40\% cross talk at their operating bias, where their definition is the ratio of
the 2-pixel peak to the 1-pixel peak in the noise response. The investigation of
the cross talk at the biases used in the test beam is in progress using several
different techniques. The results are not yet mature enough to be discussed
here. However, it should be pointed out that when using a preliminary crosstalk
correction, for example, Figure~\ref{fig:transverse} shows a much more uniform
response across the two counters. We find using these preliminary results that
photoelectron yields are roughly a factor of two less than pixel yield. On a
side note, the SiPM model tested is no longer being produced. A different device
with much lower crosstalk will be tested in a test-beam run early in 2016.

\section{Conclusions}
The Mu2e cosmic ray veto (CRV) system is an essential component of the Mu2e
experiment. The test beam measurements were carried out to understand the
performance of the scintillator counters used in the Mu2e CRV system to ensure
that they satisfy the design requirements.

The test results demonstrate a generally uniform response of the counter along
its width with a sharp drop-off $0.5~\mathrm{mm}$ away from the edges. Between
the two counters of a di-counter an effective gap of $1~\mathrm{mm}$
width is present. Fiber holes in the counters have a small effect on
the signal level when a particle passes through the hole. The response of the
counter along its length is consistent with the fiber attenuation length.


\end{document}